\documentclass[seceq]{ptptex}

\usepackage{bm}

\title{Unobservable Higgs Boson and \\
Spontaneous Violation of Lorentz Invariance}

\author{Noboru \textsc{Nakanishi}\footnote{Professor Emeritus of 
Kyoto University. \ E-mail: nbr-nak@trio.plala.or.jp}}

\inst{12-20 Asahigaoka-cho, Hirakata 573-0026, Japan}
 
\abst{The standard theory of elementary particle physics is modified in such a 
way that the Higgs boson becomes unobservable and Lorentz invariance is slightly
violated at the level of the S-matrix. The basic technique of realizing these 
properties without violating the unitarity of the physical S-matrix
is the use of the complex-ghost quantum field theory. 
}

\begin{document}

\maketitle

\section{Introduction}

The standard theory of elementary particle physics (electroweak theory plus quantum
chromodynamics) is a very successful theory.
It is formulated as a manifestly covariant quantum field theory quite 
satisfactorily$^{1)}$ and its predictions have
no clear contradictions with any high-energy experimental results,
provided that the right-handed neutrinos are taken into account.  
The only possible troublesome experimental facts against the standard theory
are the non-observation
of the Higgs boson and the observation of ultra-high energy cosmic-ray
particles indicating ``slight" \footnote{Here, ``slight" means that the violation 
may be sufficiently small for the Lorentz transformations which
are realizable by the present-day fixed-target and colliding-beam experiments.}
 violation of Lorentz invariance. 
Of course, it is quite likely that they may be resolved by 
further investigation. But there is also the possibility that these
troubles are experimentally confirmed. The purpose of the present 
paper is to propose a possible minimum modification of the standard theory
which is consistent with the absence of the Higgs boson and with  slight
violation of Lorentz invariance. 

The only role of the Higgs boson in the standard theory is to give non-zero
masses to weak bosons, leptons and quarks. Its particle contents are not only
unnecessary but also unwelcome. Indeed, its self-energy Feynman integrals are
quadratically divergent; hence its radiative mass is very sensitive to the
value of a cutoff parameter; this trouble is known as the hierarchy
problem. This problem is usually supposed to be resolved by the supersymmetric
theory (SUSY), but since no superparticles are observed yet,
it is quite unlikely that SUSY is realized in the nature.
Then it is quite difficult to resolve the hierarchy problem by means of
a renormalizable theory.

Soon after the existence of cosmic background radiation had been discovered,
it was pointed out that there exists an upper bound for the energies of 
cosmic-ray particles which can reach the earth because of their collisions
with background photons,$^{2),3)}$ provided that Lorentz invariance is assumed to
hold strictly. Though not yet confirmed, it seems that
\footnote{Akeno Giant Air Shower Array (AGASA) observed 
some extreme events whose energy involved is greater than $3 \times 10^{20}$
eV.} some cosmic-ray particles whose energies exceed the above bound have been
observed.
If this is true, it may imply that Lorentz invariance is slightly
violated in high-energy reactions.$^{4)}$

The purpose of the present paper is to propose a theory of modifying
the standard theory in such a way that there is no observable Higgs boson and that
Lorentz invariance is slightly violated. In this modification, we require
that the following properties should be kept; 1) manifest covariance of
the fundamental Lagrangian density, 2) renormalizablity of the
theory, 3) $SU(3) \times SU(2)_L \times U(1)$ gauge symmetry,
and, most importantly, 4) unitarity of the physical S-matrix.
The essential idea of realizing the above program is to make the Higgs
boson a complex ghost.

In the indefinite-metric theory,$^{5)}$
the eigenvalues of a hermitian
operator are not necessarily real; for example, those of the Hamiltonian 
can be complex in general. The states with complex energy are called 
complex ghosts. They are unphysical; they should not appear in the final state
of the physical S-matrix. Remarkably enough, it is known that
the energy conservation law forbids not only the appearance
of a single complex ghost but also that of a pair of a complex ghost and
its complex-conjugate ghost owing to relativistic kinematics (more precisely,
the probability of a pair production is of measure zero). Thus \textit{the
complex-ghost quantum field theory is consistent with the unitarity of the
physical S-matrix.}$^{6)-8)}$ Instead, it was shown that \textit{Lorentz invariance
is violated spontaneously in the Feynman diagram involving a complex-ghost-pair
intermediate state.}$^{9)}$
Because of this result, the complex-ghost quantum field theory has been regarded
as unrealistic. However, if Lorentz invariance is actually
broken in high-energy phenomena, the use of complex ghosts becomes quite 
welcome because the violation of Lorentz invariance is realized without violating the 
manifest covariance of the fundamental Lagrangian density.$^{10)}$

Our strategy is as follows. We introduce a new gauge symmetry, which is the 
local dilatation invariance, that is, 
the Weyl gauge symmetry, and indefinite-metric Higgs-like fields. 
The form of the interaction term between the Higgs field and the
indefinite-metric Higgs-like fields is uniquely determined by the requirements of
$SU(2)_L$ and local dilatation invariances and of renormalizability.
Owing to the presence of this term, it is possible that the Higgs field
and the indefinite-metric Higgs-like fields reduce to a pair of complex-ghost 
fields together with Nambu-Goldstone (NG) fields. 
Then these fields become unphysical without violating the
unitarity of the physical S-matrix.

The present paper is organized as follows. In \S2, we review the manifestly
covariant formalism of the complex-ghost quantum field theory, where we make 
an extension to the case in which the masses of the fundamental fields are not
equal. In \S3, we propose a new Higgs-sector Lagrangian density and analyze the
Higgs mechanism about it; we show that the Higgs field and the indefinite-metric
Higgs-like fields can reduce to a pair of complex-ghost fields
together with NG fields. The final section 
is devoted to discussion.

\section{Complex-ghost quantum field theory}

We consider a positive-norm scalar field $\phi_1 (x)$ and a negative-norm
scalar field $\phi_2 (x)$. Their free Lagrangian density with a 
mass-mixing term is given by
\begin{equation} 
\mathcal{L}=\frac{1}{2} \sum_{j=1}^2 (-1)^{j-1} (\partial^\mu
\phi_j \cdot \partial_\mu \phi_j - m_j^2 \phi_j^{\; 2})
-\gamma \phi_1 \phi_2,
\end{equation}
where the masses $m_1$ and $m_2$ are, in general, unequal, and a mixing
parameter $\gamma$ is positive.\footnote{If negative, change the sign of $\phi_2$.}

Field equations are
\begin{equation}
 \begin{split}
(\square + \alpha + \beta)\phi_1 + \gamma \phi_2 &= 0, \\
-(\square + \alpha - \beta)\phi_2 + \gamma \phi_1 &= 0,
 \end{split}
\end{equation}
where we set $\alpha = (m_1^2 + m_2^2)/2$ and $\beta = (m_1^2 - m_2^2)/2$.
The non-vanishing equal-time commutation relations are
\begin{equation}
[\partial_0 \phi_j (x), \, \phi_j (y)]_{x^0 = y^0} = -(-1)^{j-1} i\delta
(\bm{x} -\bm{y}).
\end{equation}
We set 
\begin{equation}
[\phi_j (x), \, \phi_k (y)] \equiv i\varDelta_{jk} (x-y)
\end{equation}
and 
\begin{equation}
\bm{\varDelta} \equiv \mathrm{matrix}(\varDelta_{jk}).
\end{equation}
We then have the following Cauchy problem:
\begin{equation}
[(\square^x + \alpha) \sigma_3 + \beta + \gamma \sigma_1) 
\bm{\varDelta} (x-y) =0
\end{equation}
together with
\begin{equation}
 \begin{split}
\bm{\varDelta}(x-y)|_{x^0 = y^0} &= 0, \\
\partial_0^x \bm{\varDelta} (x-y)|_{x^0 = y^0} &= -\sigma_3 \delta(
\bm{x}-\bm{y}),
 \end{split}
\end{equation}
where $\sigma_i$ denotes the Pauli matrix.

We solve the above Cauchy problem by diagonalizing (2.6).
Extending the analysis made previously$^{11)}$ to the unequal-mass case,
we obtain the following solution:
\begin{equation}
\bm{\varDelta} (x-y) = \frac{1}{2\sqrt{\gamma^2 -\beta^2}} \left[
\left(\sqrt{\gamma^2 -\beta^2} \sigma_3 -i\beta+i\gamma \sigma_1 \right)
\varDelta(x-y; \alpha + i\sqrt{\gamma^2 -\beta^2}) + \mathrm{c.c.} \right],
\end{equation}
where c.c. denotes complex conjugate. In (2.8), we have assumed
\begin{equation}
\gamma^2 > \beta^2;
\end{equation}
this is the important condition for the existence of complex
ghosts. The complex-mass $\varDelta$-function $\varDelta (\xi; M^2)$
with $M^2$ complex is simply defined by analytic continuation with respect
to $M^2$.

Introducing the complex-mass ``positive-energy"
\footnote{``Positive energy" means that the function considered is a
boundary value of an analytic function of $\xi^0$ from the lower-half plane.} 
 $\varDelta$-function
$\varDelta^{(+)} (\xi; M^2)$, we can give the explicit expressions for
Wightman functions $\langle 0|\phi_j (x) \phi_k (y) |0 \rangle$.
Then the Feynman propagators $\langle 0|\mathrm{T}
\phi_j (x) \phi_k (y) |0 \rangle$ are calculatedDIn matrix form, they are given by
\begin{equation}
\begin{split}
\bm{\varDelta}_F (x-y) = &\frac{1}{2\sqrt{\gamma^2 -\beta^2}} \Bigl[
\left( \sqrt{\gamma^2 -\beta^2} \sigma_3  -i\beta+i\gamma \sigma_1 \right)
\varDelta_F (x-y; \alpha + i\sqrt{\gamma^2 -\beta^2}) \\ 
+ & \left(\sqrt{\gamma^2 -\beta^2} \sigma_3 + i\beta -i\gamma \sigma_1 \right)
\varDelta_F (x-y; \alpha - i\sqrt{\gamma^2 -\beta^2}) \Bigr].
\end{split}
\end{equation}
Here the complex-mass Feynman function $\varDelta_F (\xi; M^2)$ 
with $\Re M^2 > 0$ is defined by$^{10)}$
\begin{equation}
\varDelta_F (\xi; M^2) = \frac{i}{(2\pi)^4} \int d\bm{p} \int_C
dp_0 \frac{e^{-ip\xi}}{p^2 - M^2},
\end{equation}
where the contour $C$ runs from $-\infty$ to $+\infty$
below the pole located at $p_0 = -\sqrt{M^2 + \bm{p}^2}$ 
and above the pole located at $p_0 = \sqrt{M^2 + \bm{p}^2}$.
Hence $C$ is strictly a complex contour if $\Im M^2 > 0$,
while it can be taken to be the real axis if $\Im M^2 < 0$.

We note that $i\bm{\varDelta}_F (x-y)$ is a fundamental solution to (2.6),
that is, 
\begin{equation}
[(\square^x + \alpha)\sigma_3 +\beta + \gamma \sigma_1] i\bm{\varDelta}_F (x-y)
= \delta^4 (x-y).
\end{equation}

Now, we introduce an interaction Lagrangian density and work in the 
interaction picture. The Dyson S-matrix can be defined if one employs a
\textit{Gaussian} adiabatic factor $e^{-\varepsilon^2 (x^0)^2}$.$^{10)}$
Everything goes in the same way as in the
ordinary case except for carrying out the integrations over time
variables. As mentioned above, the energy variable is inevitably complex-valued
in $\varDelta_F (\xi; M^2)$, and therefore we cannot naively take the
$\varepsilon \to 0$ limit so as to yield a $\delta$-function. We must extend the
concept of the $\delta$-function to the ``complex $\delta$-function", which
is defined in the following way.$^{12)}$

Let $\varphi (k_0)$ be a test function, which is an arbitrary
function holomorphic in an appropriate strip domain including the real axis; then
the complex $\delta$-function, $\delta_c (k_0 -E)$, for $E$ complex is defined by
\begin{equation}
\int_{-\infty}^{\infty} dk_0 \varphi (k_0) \delta_c (k_0 -E)
\equiv \frac{1}{2\pi i} \oint dk_0 \frac{\varphi (k_0)}{k_0-E},
\end{equation}
where the contour goes around $k_0 =E$ in the
anticlockwise direction. Of course, if $E$ is
real, the complex $\delta$-functon reduces to the ordinary $\delta$-function.

By using the complex $\delta$-function, we can calculate the 
$\varepsilon \to 0$ limit of the Dyson S-matrix, that is, we obtain
the momentm-space expression for it, to which the conventional Feynman
rules are applicable except for the modification of the Feynman 
$-i\varepsilon$ prescription for Feynman propagators. For example, 
the Feynman integral involving a complex-ghost-pair 
intermediate state is given by
\begin{equation}
 \int d \bm{q} \int_C dq_0
\frac{1}{(q^2 - M^2)[(p-q)^2 - M^{*2}]},
\end{equation}
where the contour $C$ runs from $-\infty$ to $+\infty$, passing below the
two poles located in the left and above the two poles located in the right.
Carrying out the integration over $q_0$, we find that there is no unitarity
cut on the real axis in the $p_0$ plane. This is a consequence
of the simple kinematical fact that the total energy, 
\begin{equation}
p_0 = \sqrt{M^2 +\bm{q}^2} + \sqrt{M^{*2} + (\bm{p}-\bm{q})^2},
\end{equation}
is \textit{not real} except for the special values of $\bm{q}$ satisfying $\bm{q}^2 =
(\bm{p} -\bm{q})^2$. This fact guarantees that
the unitarity of the physical S-matrix is not broken. 

As is easily proved$^{9)}$ and also confirmed by explicit calculation,$^{13)}$ 
however, (2.14) is \textit{not} Lorentz invariant.
That is, Lorentz invariance is spontaneously violated. 
One may suspect that this result could be avoided if the concept of
the complex $\delta$-function were not employed. This is not the case,
however. The old-fashioned (non-covariant) perturbation theory also yields
the same result.$^{10)}$ The complex $\delta$-function must be introduced
only for the \textit{manifestly covariant} perturbation theory.

\section{Modification of the standard theory}

In this section, we propose a modified version of the standard theory in which
the Higgs boson becomes unobservable.

We introduce a new gauge symmetry, which is the local dilatation invarance, that is,
the Weyl gauge symmetry.\footnote{If, instead, we introduce such a symmetry as
$U(1)$ or $SU(2)$, then the corresponding gauge field becomes a tachyon.} 
The Lagrangian density (including a gauge-fixing term) proper to
the Weyl gauge field $\tilde{A}_\mu$ is the well-known one. Our main concern is the modification of the Higgs sector.

We introduce a pair of hermitian scalar fields, denoted by
$\tilde{\varPhi} (x)$ and 
$\tilde{\varPhi}^{\star} (x)$; under the Weyl gauge transformation
they transform as $\tilde{\varPhi} (x) \to \tilde{\varPhi} (x)
e^{\varLambda (x)}$ and 
$\tilde{\varPhi}^{\star} (x) \to
\tilde{\varPhi}^{\star} (x)e^{-\varLambda (x)} $. 
The Higgs sector consists of the Higgs field $\varPhi (x)$,
which is, of course, an $SU(2)_L$-doublet non-hermitian scalar field,
and the newly introduced indefinite-metric Higgs-like fields  
$\tilde{\varPhi} (x)$ and $\tilde{\varPhi}^{\star} (x)$.
The most general expression for the Higgs-sector Lagrangian density
which is consistent with the gauge invariances and with renormalizability is
given by
\begin{equation}
\begin{split}
\mathcal{L}_{\mathrm{Higgs}} & =
(\mathcal{D}_L^{\; \mu} \varPhi)^\dagger (\mathcal{D}_{L\mu} \varPhi)
+ \mu^2 \varPhi^\dagger \varPhi -\frac{1}{2} \lambda (\varPhi^\dagger \varPhi)^2 \\
&-(\partial^\mu + \tilde{g} \tilde{A}^\mu)\tilde{\varPhi}^\star \cdot
(\partial_\mu - \tilde{g} \tilde{A}_\mu)\tilde{\varPhi}  - 
\tilde{\mu}^2 \tilde{\varPhi}^\star \tilde{\varPhi} 
+\frac{1}{2} \tilde{\lambda} (\tilde{\varPhi}^\star \tilde{\varPhi})^2
- \xi \varPhi^\dagger \varPhi \tilde{\varPhi}^\star \tilde{\varPhi},
\end{split}
\end{equation}
where $\mathcal{D}_{L\mu}$ denotes the $SU(2)_L$ covariant differentiation.
The parameters $\mu^2$, $\lambda$, $\tilde{\mu}^2$ and $\tilde{\lambda}$
are taken to be positive so as to realize the usual Higgs mechanism;
$\tilde{g}$ and $\xi$ must be nonzero. (If $\xi =0$ then
the Higgs-like fields decouple from the main part.)

As in the standard theory, $\varPhi$ has a nonvanishing vacuum expectation
value. We denote the second component
of $\langle 0 | \varPhi | 0 \rangle$ by $v/\sqrt{2}$; 
without loss of generality, we may assume for it to be positive.
Likewise, both $\tilde{\varPhi} (x)$ and 
$\tilde{\varPhi}^{\star} (x)$ have a nonvanishing vacuum expectation
value. Without loss of generaity, we may assume that
\begin{equation}
\langle 0 | \tilde{\varPhi} (x) | 0 \rangle =
\langle 0 | \tilde{\varPhi}^{\star} (x) | 0 \rangle = \frac{\tilde{v}}{\sqrt{2}} > 0.
\end{equation}
The doublet non-hermitian field $\varPhi$ is decomposed into a singlet hermitian
field $\varphi$ and a triplet hermitian field $\chi^a$ ($a=1,2,3$); the latter
is an NG field.
Likewise, we set
\begin{equation}
\tilde{\varPhi}=\frac{1}{\sqrt{2}} (\tilde{v} + \tilde{\varphi} 
+ \tilde{\chi}), \mspace{10mu}
\tilde{\varPhi}^{\star} =\frac{1}{\sqrt{2}} (\tilde{v} + \tilde{\varphi} 
- \tilde{\chi}),
\end{equation}
where $\tilde{\chi}$ is an NG field.

The NG fields become unphysical owing to the subsidiary
conditions in the BRS-invariant operator formalism of the gauge theory.$^{14)}$
There are no linear terms involving the NG fields; the quadratic terms 
involving the NG fields are absorbed into the mass terms of the gauge fields
by transforming the gauge fields.
It should be noted that \textit{the Weyl gauge field $\tilde{A}_\mu$ 
becomes a massive vector
field but not a tachyon field}. Thus, hereafter, we may concentrate our 
attention to the discussion on $\varphi$ and $\tilde{\varphi}$.
 
The Higgs potential part of (3.1) is written
\begin{equation}
\frac{1}{2}\mu^2 (v+\varphi)^2 - \frac{1}{8}\lambda (v+\varphi)^4
-\frac{1}{2}\tilde{\mu}^2 (\tilde{v}+\tilde{\varphi})^2 
+ \frac{1}{8}\tilde{\lambda} (\tilde{v}+\tilde{\varphi})^4
- \frac{1}{4}\xi (v+\varphi)^2 (\tilde{v}+\tilde{\varphi})^2.
\end{equation}
The linear terms of (3.4) must vanish, that is, we have
\begin{equation}
\begin{split}
\mu^2 v - \frac{1}{2}\lambda v^3  -\frac{1}{2}\xi v\tilde{v}^2 &= 0, \\
-\tilde{\mu}^2 \tilde{v} + \frac{1}{2}\tilde{\lambda} \tilde{v}^3
 -\frac{1}{2}\xi v^2 \tilde{v} &= 0.
\end{split}
\end{equation}
Solving (3.5), we obtain
\begin{equation}
\begin{split}
v^2 &= \frac{2(\mu^2 \tilde{\lambda} - \tilde{\mu}^2 \xi)}
{\lambda \tilde{\lambda}+ \xi^2}, \\
\tilde{v}^2 &= \frac{2(\tilde{\mu}^2 \lambda + \mu^2 \xi)}
{\lambda \tilde{\lambda}+ \xi^2}.
\end{split}
\end{equation}
Therefore, $\xi$ must satisfy the inequalities
\begin{equation}
\frac{\mu^2 \tilde{\lambda}}{\tilde{\mu}^2} >
\xi > - \frac{\tilde{\mu}^2 \lambda}{\mu^2}.
\end{equation}

The quadratic part of (3.4) is 
\begin{equation}
\left( \frac{\mu^2}{2} - \frac{3\lambda v^2}{4} -
\frac{\xi \tilde{v}^2}{4} \right) \varphi^2 +
\left(- \frac{\tilde{\mu}^2}{2} + \frac{3\tilde{\lambda} \tilde{v}^2}{4} 
- \frac{\xi v^2}{4} \right) \tilde{\varphi}^2 -
\xi v \tilde{v} \varphi \tilde{\varphi}.
\end{equation} 
In this way, we find that the free Lagrangian density for  
$\varphi$ and $\tilde{\varphi}$ becomes
\begin{equation}
\mathcal{L}_{\mathrm{Higgs}}^{\; (0)} =
\frac{1}{2}\left( \partial^\mu \varphi \cdot \partial_\mu \varphi
- \lambda v^2 \varphi^2 \right)
-\frac{1}{2}\left( \partial^\mu \tilde{\varphi} \cdot \partial_\mu 
\tilde{\varphi}
- \tilde{\lambda} \tilde{v}^2 \tilde{\varphi}^2 \right)
- \xi v \tilde{v} \varphi \tilde{\varphi},
\end{equation}
where use has been made of (3.6).

The condition (2.9) becomes
\begin{equation}
\xi v \tilde{v} > |\lambda v^2 - \tilde{\lambda} \tilde{v}^2 |.
\end{equation}
It is certainly possible to satisfy this inequality
if the values of the parameters are chosen appropriately.
Thus, $\varphi$ and $\tilde{\varphi}$ become complex-ghost fields.

\section{Discussion}

In the present paper, we have successfully proposed a modified version of the standard
theory in which the Higgs boson becomes unobservable without violating
the unitarity of the physical S-matrix. The essential technique is the 
use of the complex-ghost field theory. 

It seems that some people  dislike to use the indefinite-metric theory
in the physical context; according to them, indefinite metric is no more
than such auxiliary means as a regulator. Such assertion is, of course,
inadequate; indeed, the local quantum field theories of gauge fields and 
gravity can be formulated only in the framework of the indefinite-metric
theory. 

It is true that it is extremely
difficult to formulate the indefinite-metric quantum field theory in the
mathematically rigorous manner, but this fact does not mean that the 
correct physical theory must be the positive-metric quantum field theory.
The most fundamental principle of quantum theory is the superposition principle,
which is a linear property, while the norm positivity is a
nonlinear property. In the axiomatic quantum field theory, the
norm positivity is merely
postulated as an axiom; it is not the fact which is proved in some general context.
Although the constructive field theory showed the existence
of some nontrivial examples of the positive-metric quantum field theory,
the magnitude of the coupling constant must be restricted in general.
It is quite likely that if the coupling constant becomes larger beyond the
restriction, the norm positivity no longer remains valid. Indeed, such a phenomenon
is seen to exist in exactly solvable 2-dimensional models; furthermore,
in the Bethe-Salpeter formalism, the appearance of ghost bound states
is known to be inevitable for large values of the coupling constant.$^{15)}$
The present author believes that the use of indefinite metric is quite
natural in the framework of the Lagrangian quantum field theory.

It is expected that, in near future, the Large Hadron Collider
(LHC) experiment will clarify whether
or not the Higgs boson is really an observable particle. 
If it is not observed, our model should 
be examined more closely. The physical predictions of our model are, in addition
to the physical absence of the Higgs boson, slight violation of Lorentz invariance
and the existence of a new massive gauge boson. 

If the Higgs boson is observed, our model must be abandoned, but the complex-ghost
theory can still be used for explaining spontaneous violation of Lorentz invariance.
Furthermore, complex ghosts can be utilized as finite-mass regulators; 
for example, the quadratic divergence
of the Higgs-boson self-energy Feynman integrals 
caused by the Yukawa interactions can be removed by
introducing two pairs of \textit{bosonic} Weyl-spinor fields.

\end{document}